\documentclass[prb,twocolumn,showpacs,preprintnumbers,amsmath,amssymb,superscriptaddress]{revtex4-1}

\usepackage{graphicx}
\usepackage{dcolumn}
\usepackage{bm}

\def\Vec#1{\bm{#1}}

\begin{document}


\title{
Quasiclassical numerical method for mesoscopic superconductors: bound states in a circular $d$-wave island with a single vortex
}

\author{Yuki Nagai}
\affiliation{CCSE, Japan  Atomic Energy Agency, 5-1-5 Kashiwanoha, Kashiwa, Chiba, 277-8587, Japan}
\affiliation{CREST(JST), 4-1-8 Honcho, Kawaguchi, Saitama, 332-0012, Japan}
\author{K. Tanaka}
\affiliation{Department of Physics and Engineering Physics, University of Saskatchewan, 116 Science Place, Saskatoon, Saskatchewan, S7N 5E2 Canada}
\author{Nobuhiko Hayashi}
\affiliation{Nanoscience and Nanotechnology Research Center (N2RC), Osaka Prefecture University, 1-2 Gakuen-cho, Naka-ku, Sakai 599-8570, Japan}
\affiliation{CREST(JST), 4-1-8 Honcho, Kawaguchi, Saitama, 332-0012, Japan}


\date{\today}

\begin{abstract}
We demonstrate an efficient numerical method for obtaining unique solutions to the Eilenberger equation for a mesoscopic or nanoscale superconductor. In particular, we calculate the local density of states of a circular $d$-wave island containing a single vortex. The ``vortex shadow'' effect is found to strongly depend on the quasiparticle energy in such small systems. We show how to construct by geometry quasiparticle trajectories confined in a finite-size system with specular reflections at the boundary, and discuss the stability of the numerical solutions even in the case of vanishing order parameter as for nodal quasiparticles in a $d$-wave superconductor, or for quasiparticles passing through the vortex center with zero energy.
\end{abstract}

\maketitle

\section{Introduction}
Recently developed experimental techniques have made it possible to fabricate mesoscopic superconductors and to observe their electronic structure by scanning tunneling spectroscopy (STS).\cite{Nishio,Cren}
Due to finite-size effects, mesoscopic superconductors can exhibit properties that are significantly different from those of their analogous bulk materials. 
For example, vortex physics presents various intriguing phenomena in a mesoscopic system whose size is of the order of the coherence length or the penetration depth.
In particular, competition between the repulsive interaction among vortices, which tends towards formation of the Abrikosov vortex lattice, and quantum confinement effects results in a variety of vortex states that are unique to small systems.
The signature of giant vortices carrying multiple flux quanta\cite{Moshchalkov} and that of ``shell effects'' of multiple vortices, where vortices arrange themselves conforming to the shape of the sample, have been detected in submicron Al disks.\cite{Geim,Kanda} 
Depending on the size and shape of the system, a pair of vortex and antivortex can also be formed.\cite{Chibotaru} 
STS can directly probe the local density of states (LDOS) in such novel vortex states.

It is important to determine the phase of the superconducting order parameter in unconventional superconductors such as cuprates, heavy electron superconductors, and iron-based materials. 
One of the important characteristics of unconventional superconductivity is the possibility of the existence of Andreev bound states.\cite{Hu,Tanaka,Buchholtz}  
When there is a sign change in the order parameter in momentum space as in $d$-wave superconductors, Andreev bound states can be formed if the quasiparticle feels the sign change by specular reflection at a surface. 
Andreev bound states can also exist where the order parameter changes its sign in real space, e.g., around a vortex.
The formation of Andreev bound states is thus a key phenomenon that can reveal the fundamental nature of superconductivity.

In unconventional superconductors, 
phase-sensitive phenomena can be manifest in systems where interference effects can occur between a vortex and a surface.
In $d_{x^{2}-y^{2}}$-wave superconductors, the ``vortex shadow'' effect, which suppresses zero-energy density of states, has been found near a vortex in front of a reflecting 110-boundary.\cite{Graser}  
In chiral $p$-wave superconductors, low-energy Andreev bound states can be either suppressed or enhanced by a vortex, 
depending on its orientation with respect to the chirality of $p$-wave superconductivity.\cite{Yokoyama}  
Such phase-sensitive phenomena are expected to appear
in mesoscopic superconductors, where the effects of surfaces can be dominant.

The electronic structure of the vortex state has been studied in terms of microscopic 
mean-field theory, with which one can calculate the LDOS observable by STS.  
Such a microscopic calculation was performed by Gygi and Schl\"{u}ter,\cite{Gygi}
who evaluated the LDOS around a vortex by numerically solving 
the Bogoliubov-de Gennes (BdG) equations.
With the use of the quasiclassical theory of superconductivity,\cite{Larkin,Eilenberger} 
Hayashi {\it et al}.\cite{Hayashi} have 
reproduced the LDOS in the vortex state observed in NbSe$_2$ by STS.\cite{Hess}. 
The electronic structure around a vortex in a $d$-wave superconductor has been calculated
by Schopohl and Maki,\cite{Schopohl} using the Riccati parametrization\cite{Schopohl1} 
of the Eilenberger equation in the quasiclassical theory. 
The Riccati formalism has also been developed by Ashida {\it et al.} in the context of 
a boundary problem for superconductor-normal-metal interfaces.\cite{Ashida} 
Moreover, the Riccati formulation of the quasiclassical theory has been generalized 
for non-equilibrium superconductivity\cite{Eschrig1999,Eschrig2000} and diffusive systems.\cite{Eschrig2004,KTanakaP,KTanakaS}

Recent STS measurements have shown the direct evidence of giant vortices and multivortex configurations in nanoscale Pb islands.\cite{Nishio,Cren}
Theoretically,
the LDOS in the giant vortex state in $s$-wave mesoscopic disks has been calculated
by solving the BdG equations directly and selfconsistently.\cite{KTanaka}
Rigorously solving the BdG equations, however, has high computational demand, and most of the studies of mesoscopic vortex matter so far\cite{MelnikovVin,KopninMeso,MelnikovMeso,MelnikovMeso2} have been made within the semiclassical approximation to the BdG equations\cite{Andreev} or the quasiclassical theory.

Compared to the BdG equations, the Eilenberger equation is relatively easy to solve, 
especially by means of the Riccati parametrization.
However, in order to integrate the Riccati equations
one needs to know the initial values of the Riccati amplitudes,
namely, the boundary values in the case of a finite-size system.
Determining boundary conditions in the quasiclassical theory has indeed been a long standing issue (see Ref.~\onlinecite{Eschrig} and references therein).
The relatively short ``memory'' of the Riccati amplitudes of initial conditions has been exploited for integrating the Riccati equations for a finite-size system\cite{Amin} and for a vortex lattice.\cite{Miranovic}
The Riccati amplitudes, however, do not converge effectively for energy with vanishing imaginary part; or they may have to satisfy specific boundary conditions such as certain phase variation in a complex system. One way to deal with such a system with no bulk solution is to solve for the boundary values selfconsistently.\cite{KTanakaS}
Most generally, Eschrig\cite{Eschrig} has developed an efficient and stable numerical method for obtaining initial-value-independent solutions to the Eilenberger equation, including the spin degree of freedom and time dependence in general.

In this paper, we demonstrate how to efficiently obtain initial-value-independent solutions to the Riccati equations for a mesoscopic or nanoscale superconductor. In particular, we explicitly show in terms of the linearized BdG equations the numerical stability of the Riccati equation that allows different initial values to converge to one and the same solution. This stability that leads to a unique solution holds even for vanishing order parameter as for nodal quasiparticles in a $d$-wave superconductor, or for quasiparticles passing through the vortex center with zero energy. We also present a geometrical method for constructing quasiparticle trajectories confined in a finite-size system with specular reflections at the boundary. As an application of our technique, we calculate the LDOS in a circular $d$-wave island sustaining a single vortex. It is found that the ``vortex shadow effect'' strongly depends on the quasiparticle energy in such small systems.

The paper is organized as follows. 
In Sec.~II, we summarize the Riccati formalism of the quasiclassical theory for spin-singlet, equilibrium superconductivity, and discuss initial-value-independent solutions and the stability of the Riccati equations.
We introduce in Sec.~III our model of a circular $d_{x^{2}-y^{2}}$-wave island containing
a single vortex, and present results for this system and main conclusions in Secs.~IV and V, respectively. The general solution of a Riccati-type equation is presented in Appendix A.
In Appendix B, we discuss the stability of the Riccati equations in terms of analytical solutions of the bulk and in the vicinity of a single vortex. 
How to generate a path of integration with specular reflections at the boundary is illustrated for a circular disk in Appendix C. Throughout the paper $\hbar$ is taken to be unity.

\section{Formulation}
\subsection{Quasiclassical theory of superconductivity}
We introduce the quasiclassical Green function $\check{g}$ for a spin-singlet superconductor in equilibrium defined by 
\begin{align}
\check{g}(i \omega_{n},\Vec{r},\Vec{k}_{\rm F}) &= \left(\begin{array}{cc}
g & f\\
-\tilde{f} & -g
\end{array}\right),
\end{align}
which is a function of the Matsubara frequency $\omega_{n}$, the Fermi wave vector $\Vec{k}_{\rm F}$, and the spatial coordinate $\Vec{r}$.
The {\it check} $\check{A}$ signifies the $2 \times 2$ matrix structure in the Nambu-Gor'kov particle-hole space. 
The Eilenberger equation is the equation of motion for $\check{g}(i \omega_{n},\Vec{r},\Vec{k}_{\rm F})$,
\begin{align}
- i \Vec{v}_{\rm F}(\Vec{k}_{\rm F}) \cdot \Vec{\nabla} \check{g} &= 
\left[ 
i \tilde{\omega}_{n} \check{\tau}_{3} - \check{\Delta}(\Vec{r},\Vec{k}_{\rm F}),\, \check{g}
\right], \label{eq:eilen}
\end{align}
supplemented by the normalization condition,
\begin{align}
\check{g}^{2} &= \check{1}\,,
\end{align}
where $i \tilde{\omega}_{n} = i \omega_{n} + \Vec{v}_{\rm F} \cdot \frac{e}{c} \Vec{A}$ 
with $\Vec{A}$ a vector potential
and $\check{\tau}_{3}$ the Pauli matrix. 
The $\check{\Delta}(\Vec{r},\Vec{k}_{\rm F})$ is given by
\begin{align}
\check{\Delta}(\Vec{r},\Vec{k}_{\rm F}) &= 
\left[ 
\begin{array}{cc}
0 & \Delta(\Vec{r},\Vec{k}_{\rm F}) \\
-\Delta^{\ast}(\Vec{r},\Vec{k}_{\rm F}) & 0
\end{array}
\right]
\end{align}
in the Nambu-Gor'kov space.
Setting $i \omega_{n} = \epsilon + i \eta$, where $\eta$ is real and positive,
we have the retarded quasiclassical Green function. 

\subsection{Riccati formalism}
While several numerical methods have been developed for solving the Eilenberger equation,\cite{Thuneberg,ThunebergPRL,Klein,Ichioka,KopninJLTP} 
the Riccati parametrization is one of the most efficient and numerically stable techniques.
It can incorporate the normalization condition for the Green function automatically, and it is arguably the most versatile method that has a wide variety of application. For example, the Riccati formalism has been applied for calculation of the ac electromagnetic response of the vortex core\cite{Eschrig1999,Eschrig2009} and for a fully selfconsistent study of diffusive superconductor-normal-metal-superconductor junctions, which involve the proximity effect, multiple Andreev reflections, and non-equilibrium distribution functions.\cite{Eschrig2004,Cuevas} For both of these examples, there exists no study to date by any other technique. We will further elaborate on the numerical stability of integrating the Riccati equations in Section~\ref{subsec:riccati_stability}. The Riccati amplitudes $a$ and $b$ are introduced by writing $\check{g}$ as
\begin{align}
\check{g} &= 
\frac{- 1}{1 + a b} \left(\begin{array}{cc}
1-ab & 2 i a\\
-2 i b & -(1 - ab)
\end{array}\right). \label{eq:ab}
\end{align}
The Eilenberger equation (\ref{eq:eilen}) then reduces to a set of two decoupled differential equations of the Riccati type,
\begin{align}
\Vec{v}_{\rm F} \cdot \Vec{\nabla} a &= - 2 \tilde{\omega}_{n} a - \Delta^{\ast} a^{2} + \Delta\,, \label{eq:a}\\
\Vec{v}_{\rm F} \cdot \Vec{\nabla} b &= + 2 \tilde{\omega}_{n} b + \Delta b^{2} - \Delta^{\ast}. \label{eq:b}
\end{align}
Since these equations contain $\Vec{\nabla}$ only through $\Vec{v}_{\rm F} \cdot \Vec{\nabla}$,
they can be reduced to a one-dimensional problem on a straight line in the direction of the Fermi velocity $\Vec{v}_{\rm F}$:
\begin{align}
v_{\rm F} \frac{\partial a}{\partial s} &= - 2 \tilde{\omega}_{n} a- \Delta^{\ast} a^{2} + \Delta\,, \label{eq:as}\\
v_{\rm F}  \frac{\partial b}{\partial s} &= + 2 \tilde{\omega}_{n} b + \Delta b^{2} - \Delta^{\ast}. \label{eq:bs}
\end{align}
The local density of states (LDOS) for an isotropic Fermi surface 
as a function of quasiparticle energy $\epsilon$ 
(with respect to the Fermi level) is given by 
\begin{align}
\nu(\Vec{r},\epsilon) &=  \nu(0) \int \frac{d \Omega_{\bf k}}{4 \pi}\, {\rm Re} \left[ \frac{1 - ab}{1 + ab} \right]_{i \omega_{n} \rightarrow \epsilon + i \eta}, \label{eq:LDOS}
\end{align}
where $\nu(0)$ is the Fermi-surface density of states, 
$d\Omega_{\bf k}$ is the solid angle,
and $\eta$ is a smearing factor as due to impurity scattering.

\subsection{Initial-value-independent solution}
We now describe how initial-value-independent solutions to the Riccati equations can be obtained without specifying the initial values. This corresponds to the case of spin-singlet, equilibrium superconductivity in the general discussion in Appendix~E of Ref.~\onlinecite{Eschrig}.
Let us consider the Riccati equation (\ref{eq:as}) with complex frequency $z=i\omega_n$,
\begin{align}
v_{\rm F} \frac{\partial a}{\partial s} &=  2 i z a - \Delta^{\ast} a^{2} + \Delta\,. \label{eq:az}
\end{align}
If we can find a particular solution $a = a_{\rm P}(s)$, the general solution can be given by (see Appendix A)
\begin{align}
a(s) &= a_{\rm P}(s) + \frac{1}{- \left( \int_{s_{0}}^{s}ds^\prime A(s') e^{-K(s')} \right)e^{K(s)} + u(s_{0}) }\,,
\end{align}
with
\begin{align}
A(s) &= -\frac{\Delta^{\ast}(s)}{v_{\rm F}}\,, \\
K(s) &= \frac{2}{v_{\rm F}} \int_{s_{0}}^{s}ds^\prime \Delta^{\ast}(s')a_{\rm P}(s') - 2 i \frac{ z}{v_{\rm F}} (s-s_{0})\,. \label{eq:K}
\end{align}
The $u(s_{0})$ satisfies the initial condition at $s = s_{0}$,
\begin{align}
a(s_{0}) &= a_{\rm P}(s_{0}) + \frac{1}{u(s_{0})}.
\end{align}
If the condition
\begin{align}
\lim_{s \rightarrow \infty} K(s) = + \infty
\end{align}
is satisfied in the upper half plane of $z$, the solution $a(s)$ does not depend on $u(s_{0})$ in the limit $s \rightarrow \infty$:
\begin{align}
\lim_{s \rightarrow \infty} a(s) &= a_{\rm P}(s). 
\end{align}

Now suppose that we have obtained a numerical solution $a_{\rm N}(s)$ with the initial value at $s = s_{0}$, 
\begin{align}
a_{\rm N}(s_{0}) &= a_{0}.
\end{align}
Then another solution $a_{\rm N}'(s)$ with a different initial value $a_{0}'$ at $s = s_{0}$ can found by
\begin{align}
a_{\rm N}'(s) &= a_{\rm N}(s) + \frac{1}{- \left( \int_{s_{0}}^{s}ds^\prime A(s') e^{-K(s')} \right)e^{K(s)} + u(s_{0}) }, \label{eq:an}
\end{align}
where
\begin{align}
\frac{1}{u(s_{0})} &= a_{0}' - a_{0}\,.
\end{align}
From Eq.~(\ref{eq:K}), if
$\frac{2}{v_{\rm F}} \int_{s_{0}}^{s}ds^\prime {\rm Re} \left[ \Delta^{\ast}(s^\prime)a_{\rm N}(s^\prime)\right]$
is an increasing function of $s$ in the upper half plane of $z$, 
$e^{K(s)}$ increases with increasing $s$, since the second term in Eq.~(\ref{eq:K}) is always a monotonically increasing function in the upper half plane of $z$. 
The length is characterized by the $\Vec{k}_{\rm F}$-dependent coherence length $\xi(\Vec{k}_{\rm F}) \equiv v_{\rm F}(\Vec{k}_{\rm F})/\Delta(\Vec{k}_{\rm F})$. 
In the region $s - s_{0} \gg \xi(\Vec{k}_{\rm F})$, we have 
\begin{align}
a_{\rm N}(s) &= a_{\rm N}'(s)\,,
\end{align}
since the second term in Eq.~(\ref{eq:an}) vanishes. Thus
one can always find a numerically stable solution $a_{\rm N}(s)$
which is independent of the initial value if far away enough from the initial point. 
This stems from the fact that the numerical solution $a_{\rm N}(s)$ can be regarded as a particular solution to the differential equation (\ref{eq:as}).\cite{Eschrig}
We find that the relation ${\rm Re} \left[ \Delta^{\ast}(s)a_{\rm N}(s)\right] > 0$ is satisfied for a wide range of $s$ in various systems (see Appendix B). 
The similar argument can be made when integrating the Riccati equation for $b$ in Eq.~(\ref{eq:bs}). 

The above discussion clearly shows the reason why one has to integrate Eq.~(\ref{eq:as}) in the direction of increasing $s$ and Eq.~(\ref{eq:bs}) in the opposite direction of decreasing $s$. 
In the upper half plane of $z$, the second term in Eq.~(\ref{eq:K}) increases monotonically with increasing $s$. 
On the other hand, one has to integrate Eq.~(\ref{eq:bs}) in the direction of decreasing $s$, when considering the lower half plane of $z$.

\subsection{Choice of initial values}
\label{subsec:initial_values}
In actual calculation, one has to choose an initial value in order to numerically integrate the Riccati equation. 
We now show that $a_{0}(s_{0}) = 0$ is the best choice for the initial value for integrating Eq.~(\ref{eq:as}) {\it regardless} of the magnitude of $\Delta(s)$.
As the most extreme case, let us consider 
the quasiparticle motion for vanishing order parameter 
$\Delta(s) = 0$, e.g., for nodal quasiparticles in a $d$-wave superconductor.
The Riccati equation (\ref{eq:az}) reduces to
\begin{align}
v_{\rm F} \frac{\partial a}{\partial s} &=  2 i z a\,.
\end{align}
The solution of this linear differential equation can be expressed as 
\begin{align}
a(s) &= \exp \left[ \frac{2 i z}{v_{\rm F}} (s-s_{0})\right] a_{0}, \label{eq:a0}
\end{align}
with the initial value $a_{0}$ at $s = s_{0}$. 
This solution implies that the healing or relaxation length of the solution for $\Delta(s) \sim 0$ is roughly $v_{\rm F}/{\rm Im}z$. 
In the upper half plane of $z$ in Eq.~(\ref{eq:a0}),
\begin{align}
\lim_{s \rightarrow \infty} a(s) = 0\,.
\end{align}
This is the solution for the normal state, in which case the quasiclassical Green function $\check{g}$ is diagonal (see Eq.~(\ref{eq:ab})). 
The healing length is relatively long when $\Delta(s) \sim 0$ and ${\rm Im}z$ is small, i.e., for nodal quasiparticles.
Thus, if the initial value is much different from zero, one would need a large integration range to reach a solution $a(s) \sim 0$. In other words, the smaller the $a_{0}(s_{0})$, the shorter the integration range required.
For anti-nodal quasiparticles, the healing length is short and
the initial value hardly affects the solution so that
one can simply take $a_{0}(s_{0})$ to be zero.
Hence $a_{0}(s_{0})=0$ is the most efficient initial value for 
integrating the Riccati equation for the whole Fermi surface -- regardless of the pairing symmetry.

\subsection{Numerical stability of the Riccati equations}
\label{subsec:riccati_stability}
In this section, we show in terms of the linearized Bogoliubov-de Gennes (BdG) equations that integrating the Riccati equations is more numerically stable and effective than directly integrating the Eilenberger equation.
In general, the Riccati-type first-order nonlinear differential equation can be rewritten as a two-component first-order linear differential equation. 
For a superconducting system, these components are known as the linearized BdG equations, or the Andreev equations.\cite{Andreev,Bruder,Schopohl1}
The linearized BdG equations can be expressed as \cite{Schopohl1}
\begin{align}
v_{\rm F} \frac{\partial}{\partial s}
 \left[\begin{array}{c}
 u(s) \\
 v(s)
\end{array}\right] &= 
\hat{K} 
 \left[\begin{array}{c}
 u(s) \\
 v(s)
\end{array}\right], \label{eq:lBdG}
\end{align}
where
\begin{align}
\hat{K}&\equiv \left[\begin{array}{cc}- \tilde{\omega}_{n} & - i \Delta(s) \\
i \Delta^{\dagger}(s) & \tilde{\omega}_{n}\end{array}\right].
\end{align}
The Riccati equation (\ref{eq:a}) can be derived by defining
\begin{align}
a(s) &= i \frac{u(s)}{v(s)} \label{eq:auv}
\end{align}
in the linearized BdG equations (\ref{eq:lBdG}).

To obtain a formal solution, we assume $\Delta(s)$ to be a piecewise-constant function. 
In the interval $s_{i} < s < s_{i+1}$, where $\hat{K}$ is constant, 
one can solve Eq.~(\ref{eq:lBdG}) as 
\begin{align}
 \left[\begin{array}{c}
 u(s) \\
 v(s)
\end{array}\right] &= 
\exp \left[ \hat{K}' \right] 
 \left[\begin{array}{c}
 u_{0} \\
 v_{0}
\end{array}\right], 
\end{align}
with 
\begin{align}
\hat{K}' &\equiv 
\frac{1}{v_{\rm F}}\hat{K}(s - s_{i})
= \hat{U} \left[\begin{array}{cc}E(s) &0 \\
0& -E(s)\end{array}\right] \hat{U}^{\dagger}.
\end{align}
The eigenvalues are given by
\begin{align}
E(s) &= \sqrt{A^{2} + |B|^{2}} (s - s_{i}),\label{es} \\
A &\equiv  \frac{1}{v_{\rm F}}\tilde{\omega}_{n}, \\
B &\equiv -\frac{i}{v_{\rm F}} \Delta,
\end{align}
and the unitary matrix $\hat{U}$ can be written as
\begin{align}
\hat{U} &= \left[\begin{array}{cc}\alpha &-\beta^{\ast} \\
\beta&\alpha^{\ast}\end{array}\right].
\end{align}
Starting from the initial values $(u_{0},v_{0})$ at $s=s_{0}$, 
one can construct a solution $(u,v)$ for the entire space by connecting the solutions at each boundary between two adjacent piece-wise regions.
We note that $E(s)$ is positive and  increases monotonically with $s$ even when $\Delta = 0$, if one considers the upper half plane of complex frequency.
Thus the healing or relaxation length of the solution is determined by 
whichever between $A$ and $B$ has the shortest characteristic length scale.
This in fact guarantees that 
the healing length does not diverge even for a quasiparticle moving along the nodal direction of the order parameter, where $\Delta = 0$, because of the $A$ term in Eq.~(\ref{es}).

The general solution can be written as
\begin{align}
u(s) &= \alpha(s) e^{E (s)} u'(s)- \beta^{\ast}(s) e^{- E(s)} v'(s), \label{eq:uu}\\
v(s) &= \beta(s) e^{E(s)} u'(s)+ \alpha^{\ast}(s) e^{- E(s)} v'(s), \label{eq:vv}
\end{align}
where
\begin{align}
 \left[\begin{array}{c}
 u'(s) \\
 v'(s)
\end{array}\right] &= 
\hat{U}^{\dagger}(s)  \left[\begin{array}{c}
 u_{0} \\
 v_{0}
\end{array}\right].
\end{align}
If one wants to integrate the linearized BdG equations in the direction of increasing $s$, one has to carefully choose the initial condition $(u_{0}, v_{0})$ to avoid divergence in the limit $s \rightarrow \infty$.  

On the contrary, the solutions $a(s)$ and $b(s)$ of the Riccati equations are 
numerically stable and one needs not worry about divergence. 
This can be seen by substituting Eqs.~(\ref{eq:uu}) and (\ref{eq:vv}) into
Eq.~(\ref{eq:auv}):
\begin{align}
a(s) &= i\frac{ \alpha(s) e^{E(s)} u'(s)- \beta^{\ast}(s) e^{- E(s)} v'(s)}
{\beta(s) e^{E(s)} u'(s)+ \alpha^{\ast}(s) e^{- E(s)} v'(s)}.
\end{align}
In the limit $s \rightarrow \infty$,  
$a(s) \rightarrow i\alpha(s)/\beta(s)$
and thus $a(s)$ never diverges in this limit.
Furthermore, it is evident from the linearized BdG equations that
initial-value-independent solutions can be obtained for $\alpha(s)$ and $\beta(s)$,
which contain neither $u_{0}$ nor $v_{0}$ in the limit $s \rightarrow  \infty$. 
It should be noted, however, that $\alpha(s)$ and $\beta(s)$ depend on the initial coordinate $s_{0}$ as so do $A(s)$ and $B(s)$.

The solutions to the linearized BdG equations in Eqs.~(\ref{eq:uu}) and (\ref{eq:vv}) are a linear superposition of two unbounded solutions with the factors $e^{E(s)}$ and $e^{-E(s)}$.
The Eilenberger equation also has diverging or ``exploding'' solutions.\cite{Thuneberg} 
The so-called explosion method is based on the fact that a bounded solution 
to the Eilenberger equation can be constructed using the commutator 
of two unbounded solutions.\cite{ThunebergPRL}
Since unbounded solutions are numerically unstable and the method relies on cancellation of large numbers, a careful computational treatment is required for integrating the linearized BdG equations or the Eilenberger equation using the explosion method.

In contrast, physical, bounded solutions can be constructed for the Riccati equations
without any difficulty owing to unphysical, unbounded solutions.\cite{ThunebergPRL}
The $a(s)$, which consists of exploding solutions $u(s)$ and $v(s)$ with the factor 
$e^{E(s)}$, can be obtained by simply integrating Eq.~(\ref{eq:as}) in the direction of increasing $s$. Similarly, Eq.~(\ref{eq:bs}) can be integrated in the opposite direction to find $b(s)$, which consists of other exploding solutions $u(s)$ and $v(s)$ with the factor $e^{-E(s)}$. One can then construct a physical, bounded quasiclassical Green function $\check{g}$ from Eq.~(\ref{eq:ab}).
Hence, the Riccati parametrization makes solving the Eilenberger equation for the quasiclassical Green function more numerically stable and effective.

The stability of the Riccati equations as demonstrated above can also be shown for the more general case of spin- and time-dependent superconductivity.\cite{Matthias}

\section{Numerical method}
We illustrate how initial-value-independent solutions can be obtained
for a circular $d$-wave island containing a single vortex.
The method of generating paths described in this section can be generalized for a finite-size system of any shape with specular reflections at the boundary.

\subsection{Model}
We consider a two-dimensional system of circular shape of radius $r_{c}$,
which has a specular surface and a circular Fermi surface.
The boundary condition can then be expressed as\cite{Graser} 
\begin{align}
a(|\Vec{r}| = r_{c},\Vec{k}_{\rm in}) &= a(|\Vec{r}| = r_{c},\Vec{k}_{\rm out})\,, \\
b(|\Vec{r}| = r_{c},\Vec{k}_{\rm in}) &= b(|\Vec{r}| = r_{c},\Vec{k}_{\rm out})\,.
\end{align}
Here $\Vec{k}_{\rm in}$ is connected with $\Vec{k}_{\rm out}$ by specular reflection.
We introduce a pairing potential of the form,
\begin{align}
\Delta(\Vec{r},\Vec{k}_{\rm F}) &= \Delta_{0} f(r) d(\Vec{k}_{\rm F}) e^{i \alpha},
\end{align}
where $\Vec{r}=r(\cos\alpha,\sin\alpha)$ in polar coordinates.
Here $f(r)$ gives the spatial (radial) variation of the pairing potential
with $f(0) = 0$ at the vortex center,
and $d(\Vec{k}_{\rm F})$ describes the gap anisotropy in momentum space. 
The direction of the quasiparticle motion is characterized by angle $\theta$ 
in two-dimensional momentum space. 
We consider a $d_{x^{2}-y^{2}}$-wave superconductor with
\begin{align}
d(\Vec{k}_{\rm F}) &= \cos 2 \theta.
\end{align}
Considering the strongly type-II limit, we neglect the vector potential: $i \tilde{\omega}_{n} \rightarrow i \omega_{n}$. 
Setting $i \omega_{n} \rightarrow \epsilon + i \eta$, we integrate the Riccati equations by means of the fourth-order Runge-Kutta method.
For the sake of illustrating the numerical technique, we present results for a given paring potential with $f(r) = r/\sqrt{r^{2} + \xi_{0}^{2}}$
and use the length unit $\xi_{0} \equiv v_{\rm F}/\Delta_{0}$ and smearing factor
$\eta = 0.01\Delta_{0}$.
To obtain $a(x,y,\theta)$ and $b(x,y,\theta)$, we must integrate the Riccati equations along paths that are specularly reflected at the boundary and thus confined within the circle.

\subsection{Numerical recipe}
\begin{figure}[thb]
\includegraphics[width = 5cm]{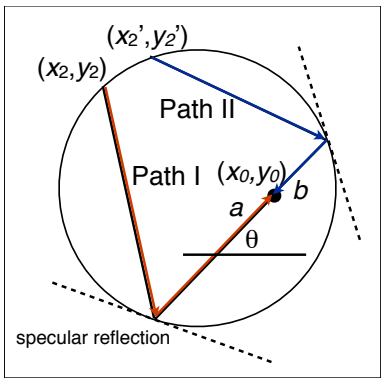}
\caption{\label{fig:cir}
(Color online)
Schematic plot of quasiparticle paths with multiple specular reflections.
}
\end{figure}

Let us now describe how to obtain the Riccati amplitudes $a$ and $b$ at point $(x_{0},y_{0})$ for a given momentum direction $\theta$. Starting from the point of interest $(x_{0},y_{0})$, we first generate path I for $a$ as indicated in Fig.~\ref{fig:cir} by drawing the path in the opposite direction (i.e., in the direction of $\theta+\pi$) to the point of the $n$-th specular reflection $(x_{n},y_{n})$ at the boundary (See Appendix C). Similarly, we generate path II for $b$ in the opposite direction from $(x_{0},y_{0})$ to $(x_{n}^\prime,y_{n}^\prime)$ with $n$ reflections.  The length of the paths after $n$ reflections should be much longer than the coherence length; i.e., the smaller the system size, the larger the $n$ should be.

Next we integrate the Riccati equations for $a$ and $b$ from $(x_{n},y_{n})$ and 
$(x_{n}',y_{n}')$, respectively, to $(x_{0},y_{0})$ for momentum direction $\theta$.
With the use of Eq.~(\ref{eq:LDOS}) we can obtain the LDOS of a circular $d$-wave island.
One must make sure that 
the results do not depend on the initial value nor the length of the path.   
\section{Results}
We present the LDOS with and without a vortex in a circular $d$-wave island.
\subsection{Without a vortex}
\begin{figure}[tbh]
  \begin{center}
    \begin{tabular}{p{0.5 \columnwidth} p{0.5 \columnwidth}}
      \resizebox{0.5 \columnwidth}{!}{{\Huge (a) \: } \includegraphics{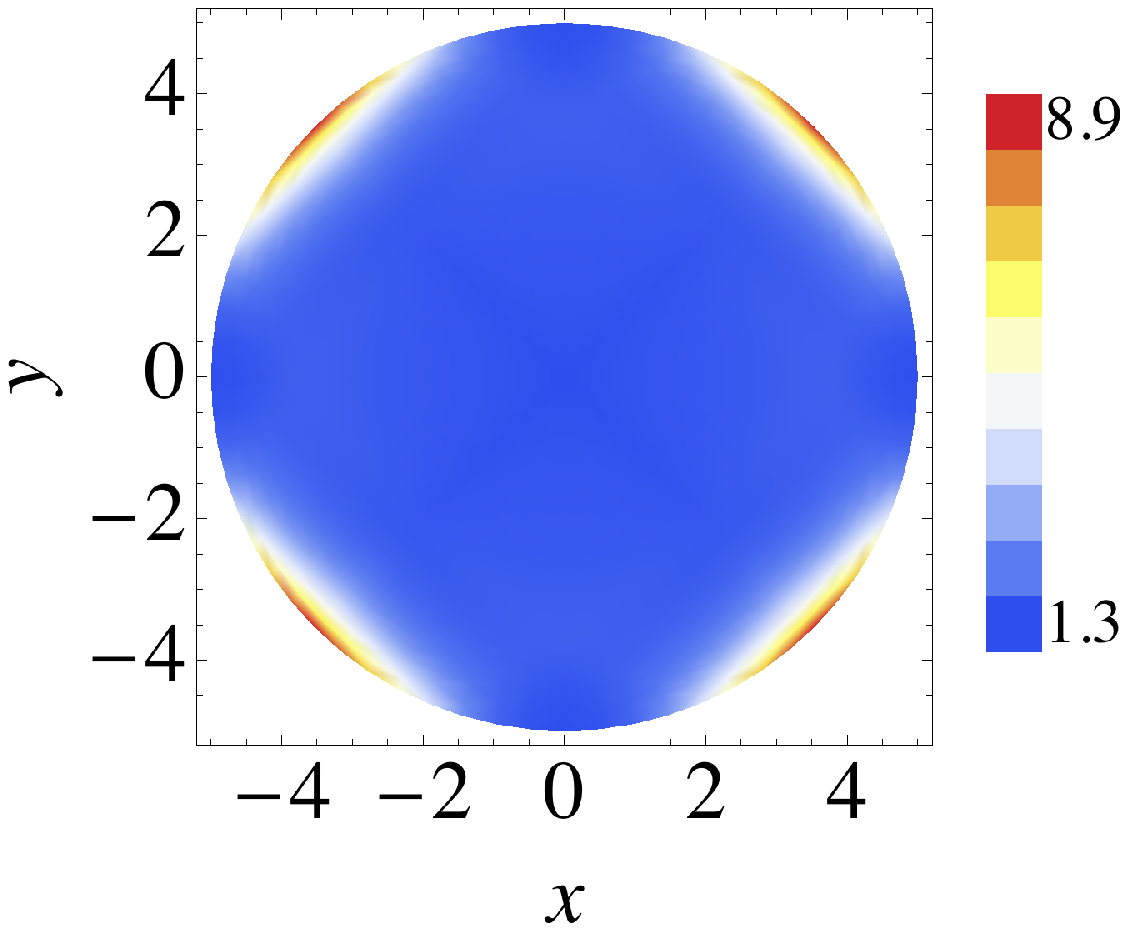}} &
      \resizebox{0.5 \columnwidth}{!}{{\Huge (b) \: } \includegraphics{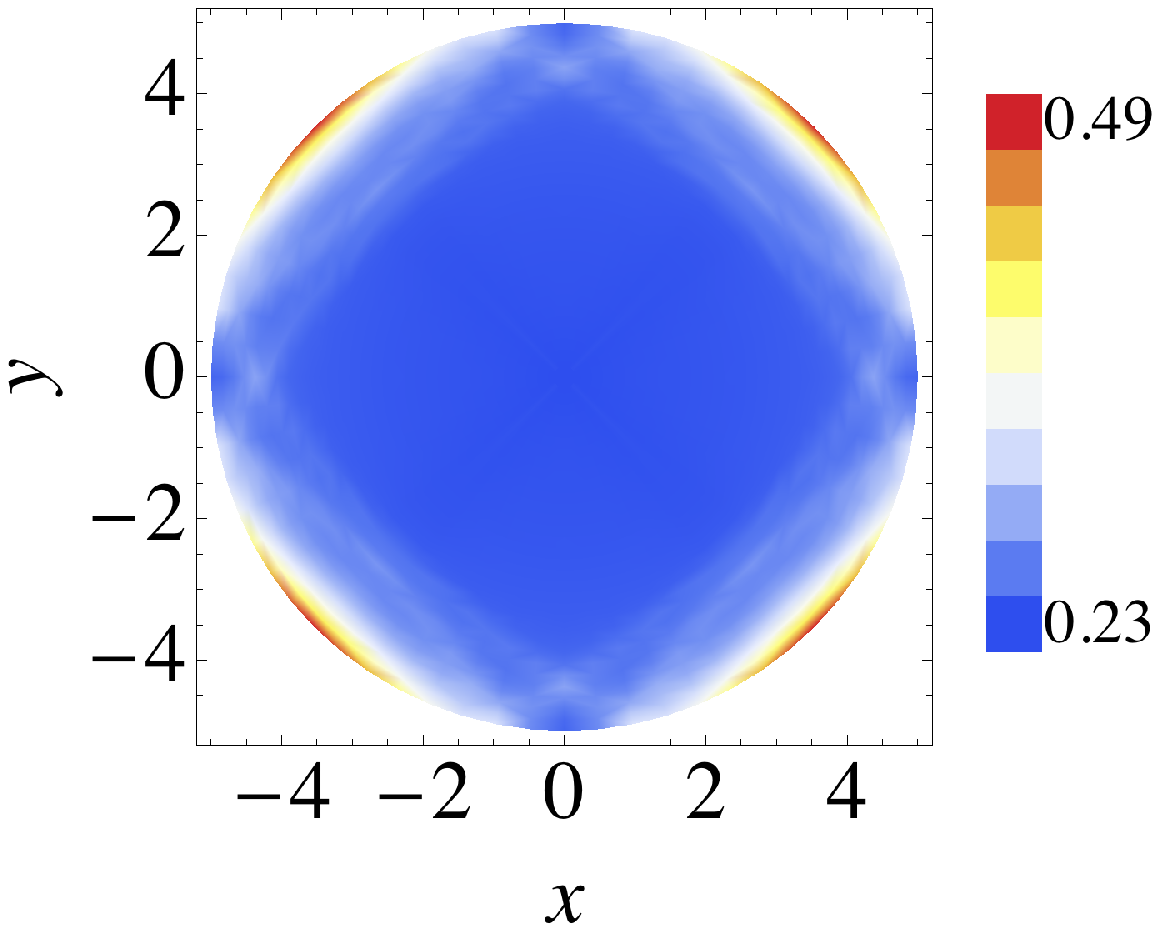}} 
    \end{tabular}
    \resizebox{0.5 \columnwidth}{!}{{\Huge (c) \: } \includegraphics{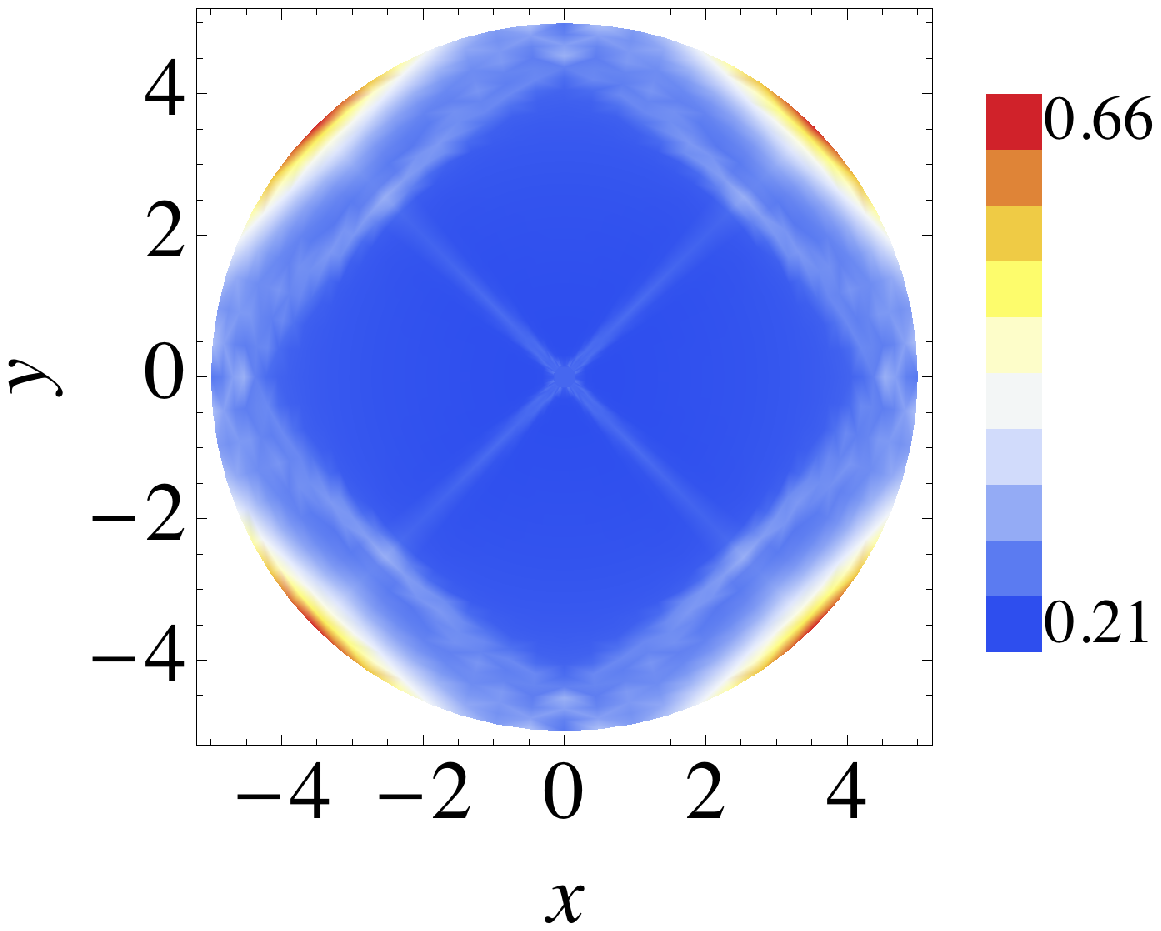}} 
\caption{\label{fig:nov}
(Color online) 
Local density of states of a circular $d_{x^{2}-y^{2}}$-wave island without a vortex 
for energy (a) $\epsilon = 0$, (b) $0.05 \Delta_{0}$, and (c) $0.1\Delta_{0}$. 
The radius is $r_{c} = 5 \xi_{0}$ and the smearing factor $\eta = 0.01\Delta_{0}$.
}
  \end{center}
\end{figure}

First we discuss the LDOS for a system without a vortex.
Shown in Fig.~\ref{fig:nov} is the LDOS given by Eq.~(\ref{eq:LDOS}), 
$\nu(\Vec{r},\epsilon)$, in units of the Fermi surface density of states
for $r_{c} = 5 \xi_{0}$; for
(a) $\epsilon = 0$, (b) $0.05 \Delta_{0}$, and (c) $0.1\Delta_{0}$. 
We have used $720$ $\theta$-meshes in momentum space.
The Andreev bound states at the 110-boundaries\cite{Hu} can be seen clearly
(note that the LDOS is plotted in different scales for the different values of 
$\epsilon$).
Due to the small size, the zero-energy LDOS is nonzero over the entire system,
while the four nodal directions are visible for $\epsilon=0.1\Delta_{0}$. 
\subsection{With a single vortex}
\begin{figure}[tbh]
  \begin{center}
%
  
    \begin{tabular}{p{0.5 \columnwidth} p{0.5 \columnwidth}}
      \resizebox{0.5 \columnwidth}{!}{{\Huge (a) \: } \includegraphics{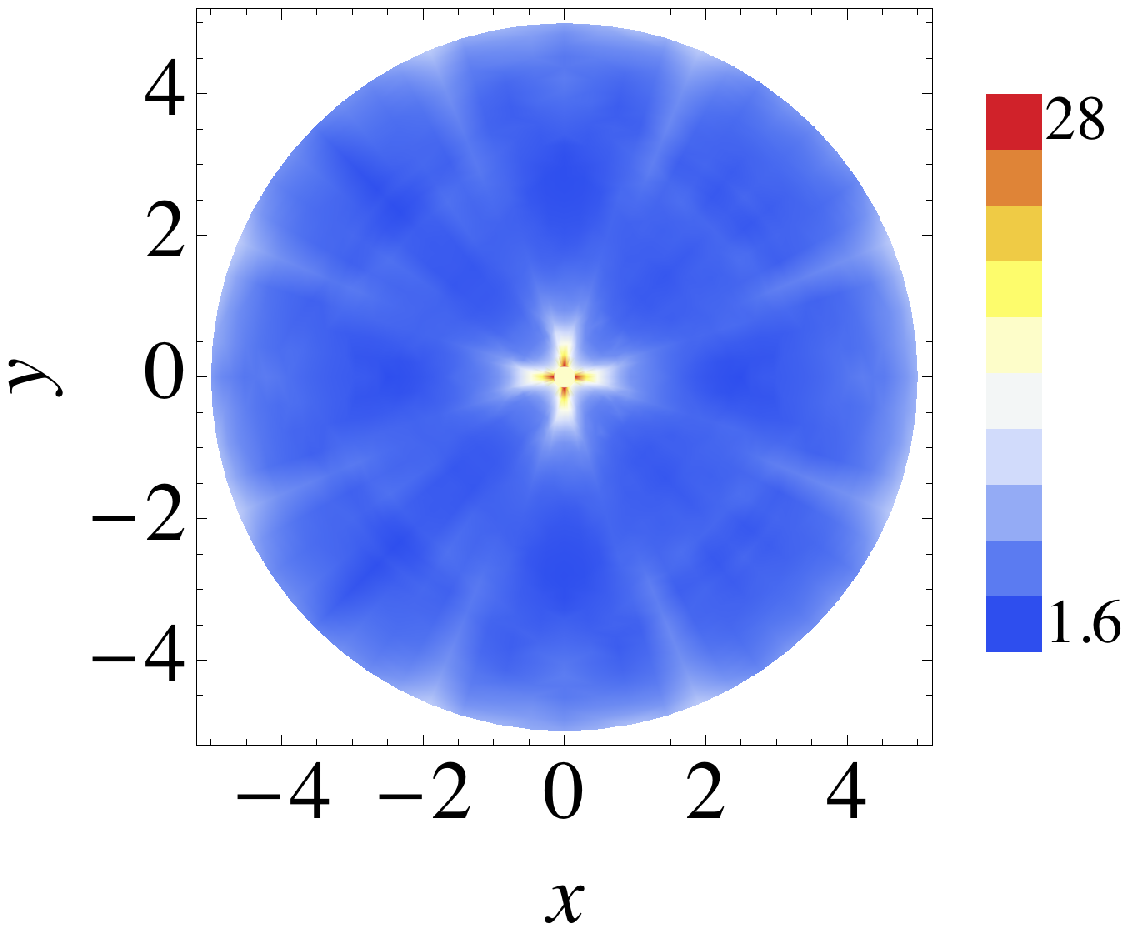}} &
      \resizebox{0.5 \columnwidth}{!}{{\Huge (b) \: } \includegraphics{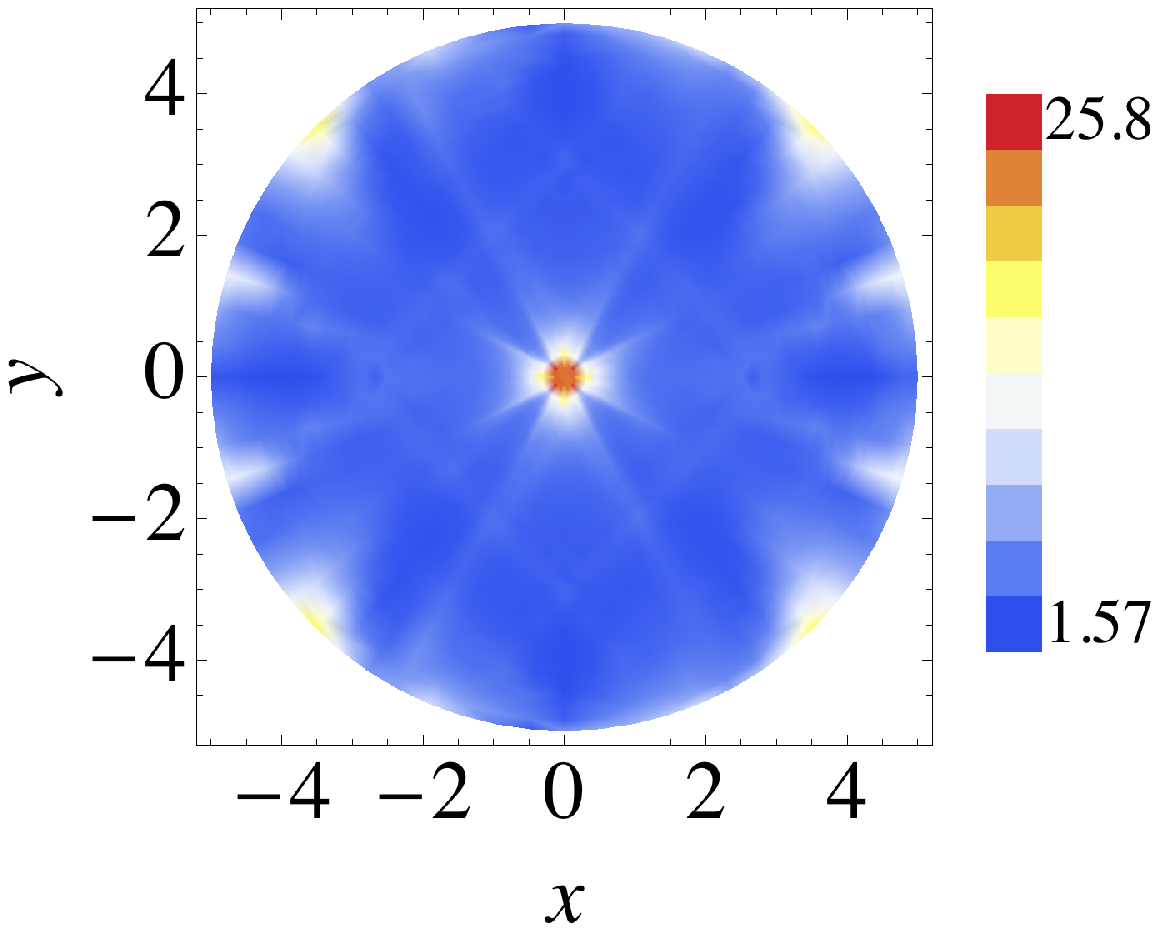}} 
    \end{tabular}
    \resizebox{0.5 \columnwidth}{!}{{\Huge (c) \: } \includegraphics{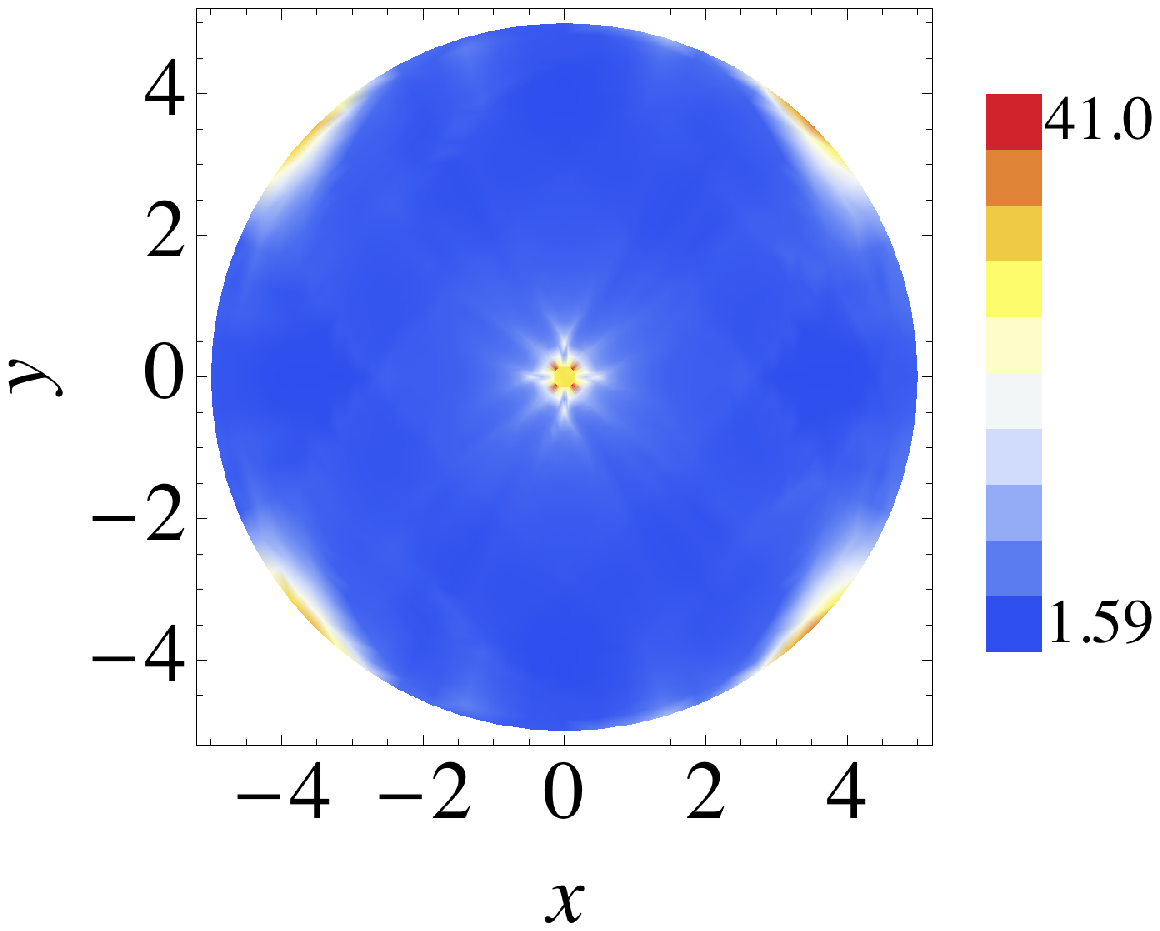}} 
\caption{\label{fig:v}
(Color online) 
Local density of states in a circular $d_{x^{2}-y^{2}}$-wave island with a vortex 
at the center for energy 
(a) $\epsilon = 0$, (b) $0.05 \Delta_{0}$, and (c) $0.1\Delta_{0}$. 
The radius is $r_{c} = 5 \xi_{0}$ and the smearing factor $\eta = 0.01\Delta_{0}$.
}
  \end{center}
\end{figure}

Next we consider a single vortex at the center of a circular $d$-wave island
with $r_{c} = 5 \xi_{0}$.
The LDOS is presented in Fig.~\ref{fig:v} for (a) $\epsilon = 0$, (b) $0.05 \Delta_{0}$, and (c) $0.1\Delta_{0}$.
The ``vortex shadow'' effect, which has been discussed by 
Graser {\it et al.}\cite{Graser} 
for a vortex near a surface of a $d$-wave superconductor, is manifest in our results.
As can be seen in Fig.~\ref{fig:v}(a), the vortex shadow effect causes zero-energy bound states to disappear.
The Andreev bound states at the 110-surfaces exist with nonzero energy, 
and their pattern changes with increasing energy.
A trajectory in the region where the LDOS becomes larger near the vortex center can be regarded as a ``ray of light'' for the surface bound states.
This can be seen clearly in Figs.~\ref{fig:v}(b) and \ref{fig:v}(c). 

\begin{figure}[htb]
  \begin{center}
%
  
      \resizebox{0.8\columnwidth}{!}{ \includegraphics{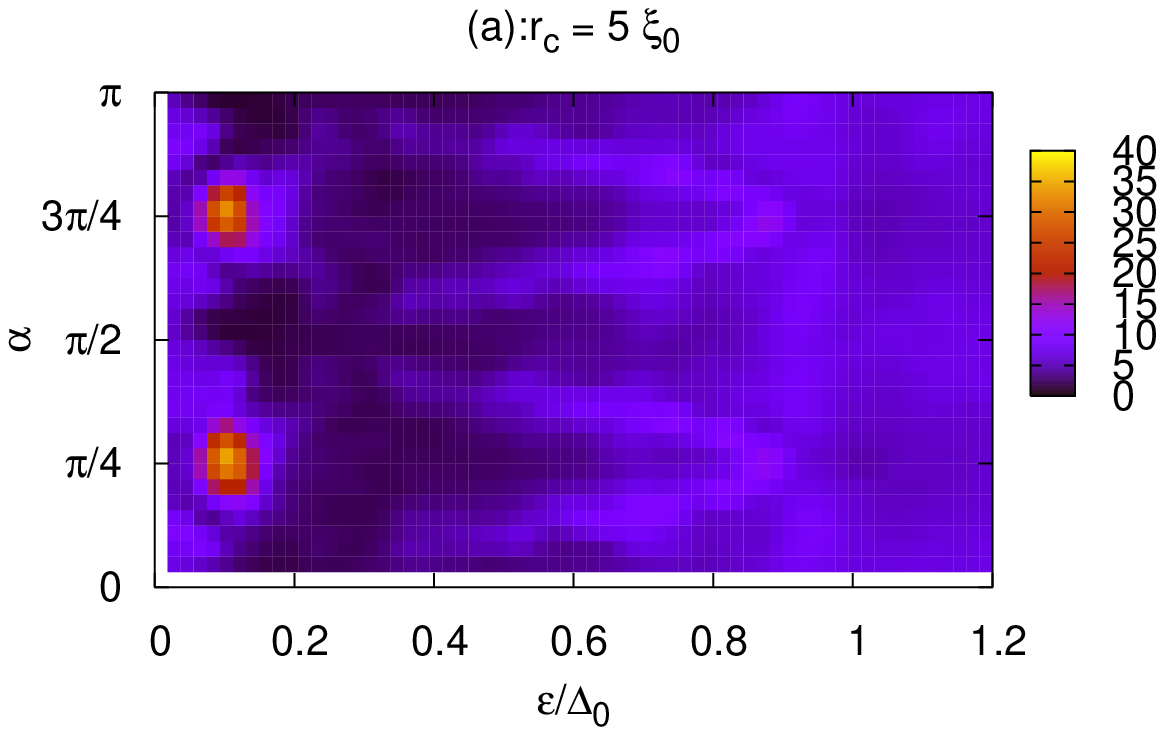}} \\
      \resizebox{0.8 \columnwidth}{!}{ \includegraphics{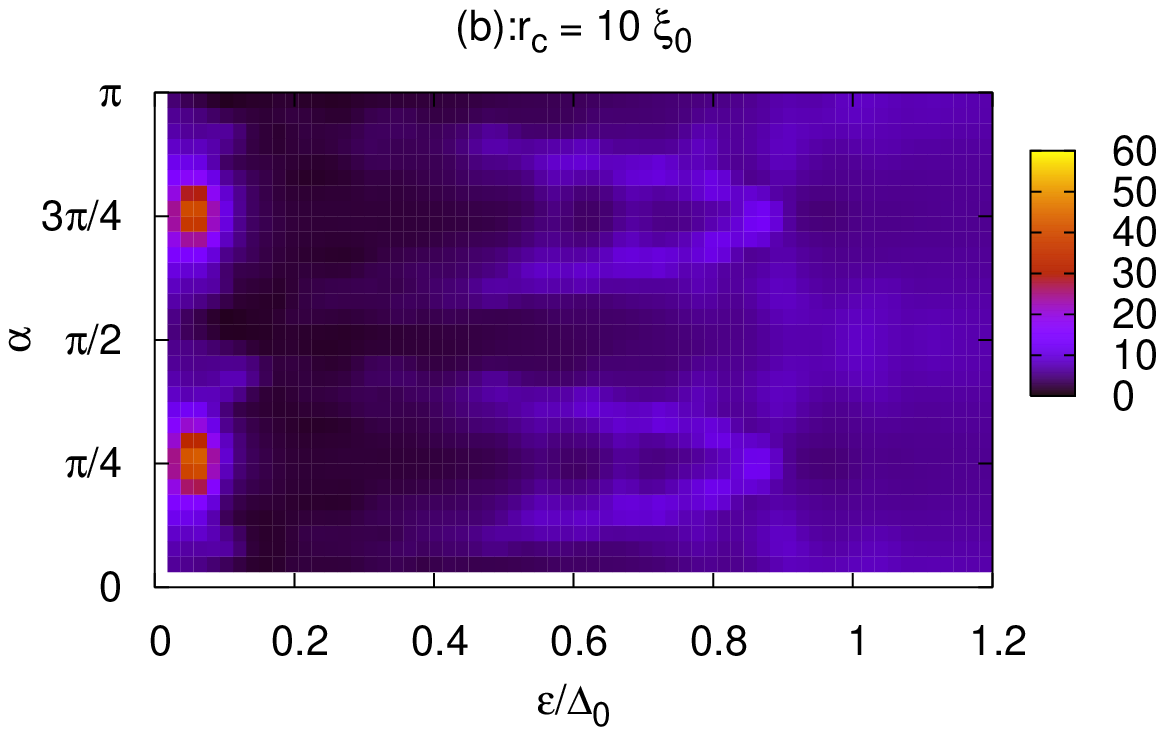}} \\
    \resizebox{0.8 \columnwidth}{!}{\includegraphics{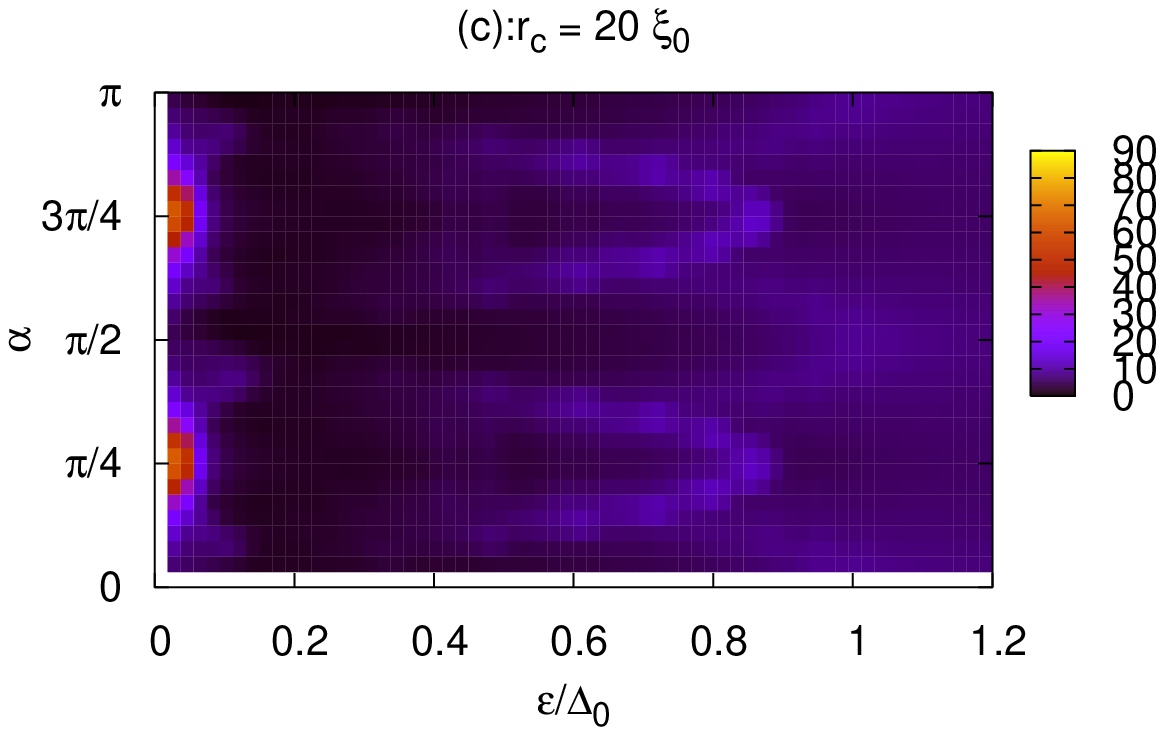}} 
\caption{\label{fig:rc}
(Color online) 
Local density of states in a circular $d_{x^{2}-y^{2}}$-wave island with a vortex 
at the center along the boundary $\Vec{r} = r_{c}(\cos \alpha,\sin \alpha)$
as a function of polar angle $\alpha$ and energy $\epsilon$, 
for radius (a) $r_{c} = 5 \xi_{0}$, (b) $10\xi_{0}$, and (c) $20\xi_{0}$.
}
  \end{center}
\end{figure}

Figure~\ref{fig:rc} illustrates the size dependence of the LDOS as a function of energy along the circumference of the system
for (a) $r_{c} = 5 \xi_{0}$, (b) $10\xi_{0}$, and (c) $20\xi_{0}$.
The vortex shadow effect diminishes as the system size increases, and 
it is indiscernible in the LDOS for $r_{c}=20\xi_{0}$ shown in Fig.~\ref{fig:rc}(c).
The $r_{c}$-dependence of our LDOS is consistent with Figure 5 of
Ref.~\onlinecite{Graser}, where the LDOS at a 110-boundary is plotted as a function of
energy for various distances of the vortex from the boundary.
 
\subsection{Demonstration of the numerical stability}
\begin{figure}[t]
  \begin{center}
    \begin{tabular}{p{\columnwidth}}
      \begin{center}
      \resizebox{0.75\columnwidth}{!}{\includegraphics{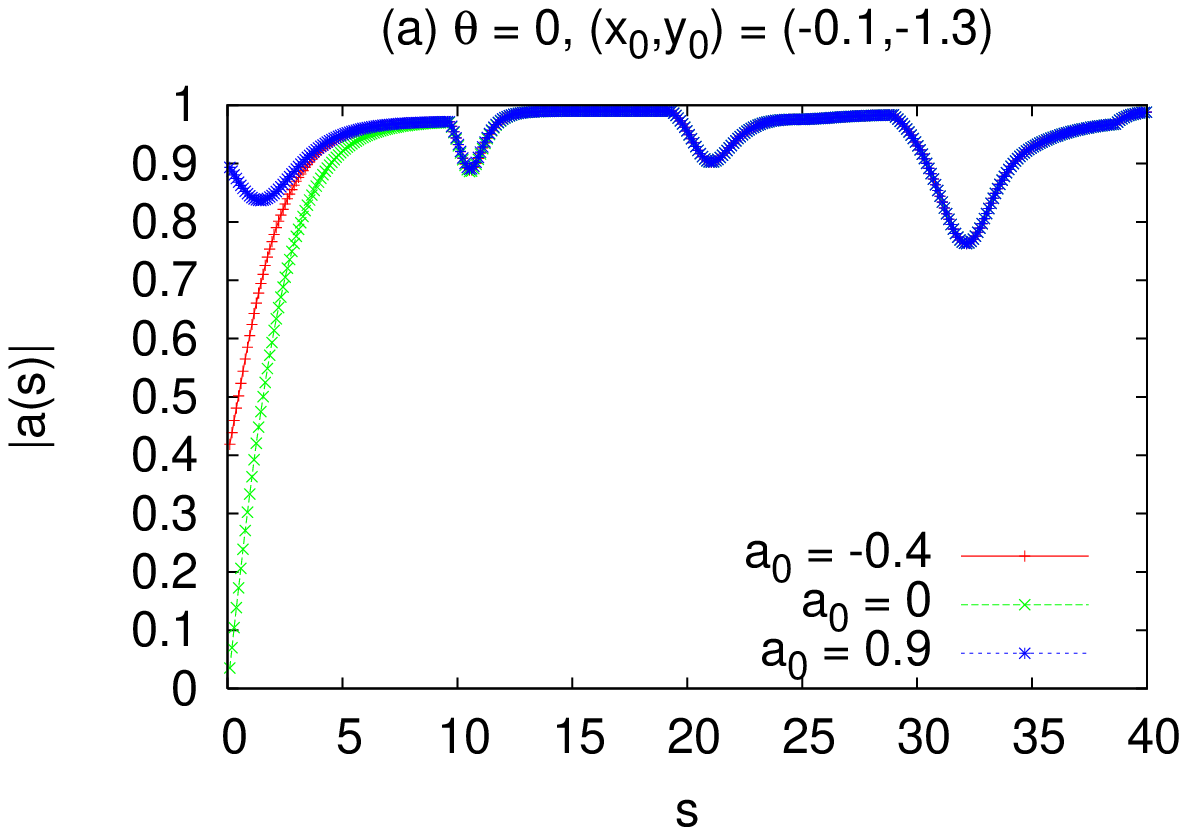}} \\
      \resizebox{0.75 \columnwidth}{!}{\includegraphics{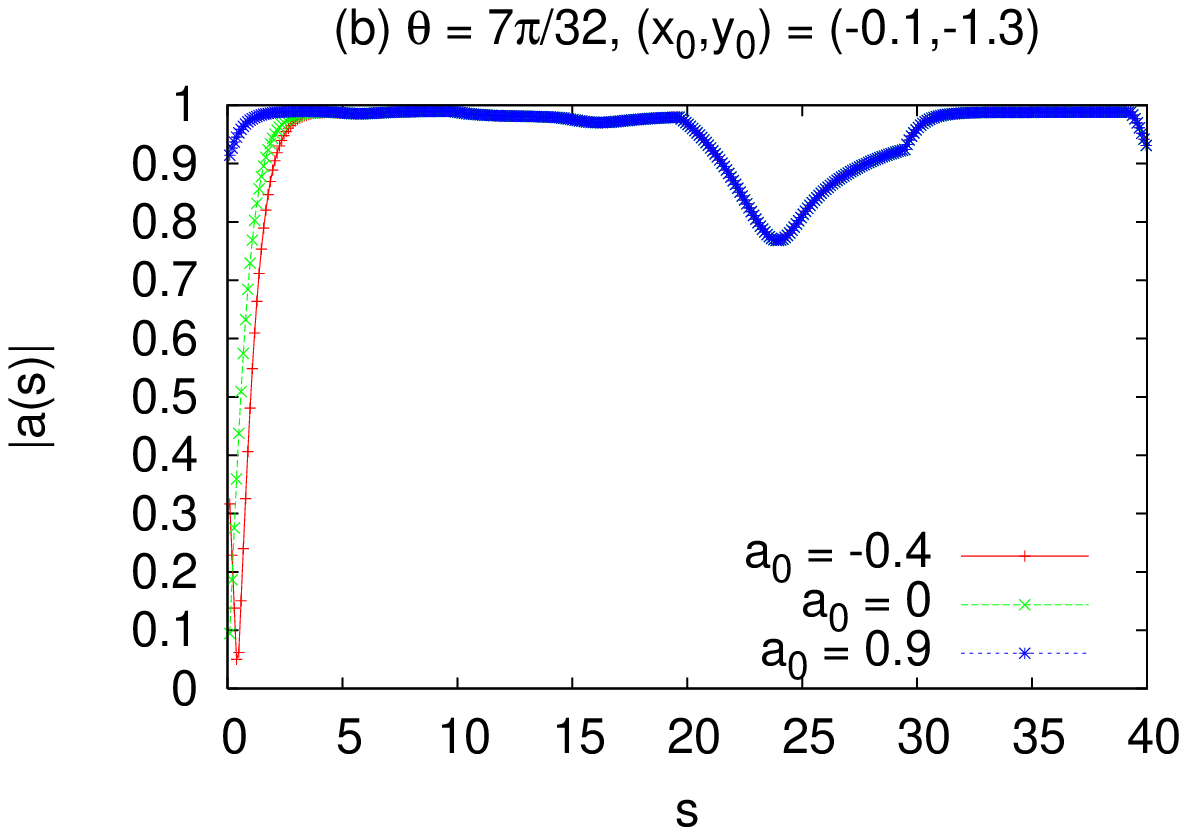}} 
      \end{center}
    \end{tabular}
\caption{\label{fig:adep}
(Color online) 
The initial-value dependence of $|a(s)|$ for momentum direction (a) $\theta = 0$ ($d(\Vec{k}_{\rm F}) = 1$) and (b) $\theta = 7\pi/32$ ($d(\Vec{k}_{\rm F}) \sim 0.2$)
as a function of integration length $s$ (in units of $\xi_0$) for a circular $d_{x^{2}-y^{2}}$-wave island. The radius is $r_{c} = 5\xi_0$ and the smearing factor $\eta = 0.01 \Delta_0$. 
}
  \end{center}
\end{figure}

Figure~\ref{fig:adep} demonstrates how quickly different initial values converge to the same solution for zero energy with the smearing factor $\eta = 0.01\Delta_{0}$. 
The upper and lower graphs in Fig.~\ref{fig:adep} show the Riccati amplitude $|a(s)|$ as a function of distance $s$ for various initial values $a_0$,
for momentum direction (a) $\theta=0$ ($d(\Vec{k}_{\rm F}) = 1$) and (b) $\theta = 7\pi/32$ ($d(\Vec{k}_{\rm F}) \sim 0.2$), respectively.
The point of interest $(x_0,y_0)=(-0.1,-1.3)$, and the initial point of integration $(x_n,y_n)$ at the boundary has been determined so that the integration path is $300 \xi_0$ or longer.
In Fig.~\ref{fig:adep} $|a(s)|$ is shown over the length $40 \xi_0$ from the initial point of integration $(x_n,y_n)$.
Although the healing length depends on the spacial variation of $\Delta(s)$ 
along the path, a converged solution can be obtained regardless of $a_0$ typically 
within a few to $\sim 10$ times the coherence length, including $a_{0}=0$ as mentioned above.

\begin{figure}[t]
  \begin{center}
    \begin{tabular}{p{\columnwidth}}
      \begin{center}
      \resizebox{\columnwidth}{!}{\includegraphics{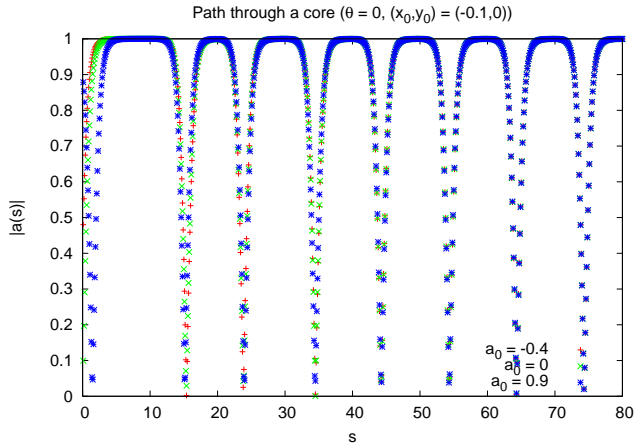}} 
      \end{center}
    \end{tabular}
\caption{\label{fig:adepzero}
(Color online) 
The initial-value dependence of $|a(s)|$ for momentum direction $\theta = 0$ ($d(\Vec{k}_{\rm F}) = 1$) as a function of integration length $s$ (in units of $\xi_0$) for a circular $d_{x^{2}-y^{2}}$-wave island, where the trajectories pass through the vortex center ($r = 0$). 
The radius is $r_{c} = 5\xi_0$ and the smearing factor $\eta = 1 \times 10^{-16} \Delta_0$. 
}
  \end{center}
\end{figure}

As discussed in Section~\ref{subsec:riccati_stability}, 
the numerical stability of the Riccati equations is indifferent to vanishing order parameter.
We have confirmed this by obtaining numerically stable solutions
in the case of nodal quasiparticles, for 
$\theta = 255\pi/1024$ ($d(\Vec{k}_{\rm F}) \sim 6 \times 10^{-3}$). 
Figure~\ref{fig:adepzero} shows that different initial values converge to a single solution even if the trajectories pass through the vortex center with zero energy and a negligible smearing factor $\eta = 1 \times 10^{-16}\Delta_{0}$. 
As the energy is so small and the order parameter vanishes at the vortex core, it takes more distance for convergence to occur in such a case, as is apparent by comparing Figs.~\ref{fig:adep} and \ref{fig:adepzero}. It can be seen in Fig.~\ref{fig:adepzero}, however, that the solution is well converged within the distance $\sim 80 \xi_0$, and hence starting the integration $300 \xi_0$ away from $(x_0,y_0)$ is sufficient even in this case.

It is also possible to find stable, unique solutions to the Riccati equations for an array of randomly distributed vortices, i.e., without any symmetry or periodicity, such as multiple vortices in a nanoscale island of arbitrary shape -- as observed in the recent experiments.\cite{Nishio,Cren}
\section{Conclusion}
In summary, we have demonstrated a numerical procedure of efficiently obtaining stable, initial-value-independent solutions to the Riccati equations for spin-singlet, equilibrium superconductivity in a finite-size system. In particular, we have shown the stability of the Riccati equations that allows one to find unique solutions in terms of the linearized Bogoliubov-de Gennes equations, and how to construct by geometry paths of integration confined by specular reflections at the boundary. We have applied this technique for calculating the local density of states in a circular $d$-wave island with a single vortex. We find that the ``vortex shadow'' effect strongly depends on the quasiparticle energy in mesoscopic or nanoscale superconductors. For the purpose of illustration, we have assumed a certain spatial variation of the order parameter. It is straightforward, however, to incorporate selfconsistency as well as to include a vector potential in this method.

\section*{Acknowledgment}
We thank M. Machida, M. Ichioka and D. A. Takahashi for helpful discussions and comments. 
The calculations have been performed using the supercomputing system 
PRIMERGY BX900 at the Japan Atomic Energy Agency. 
The research was supported partially by the Natural Sciences and Engineering Research Council of Canada.
\appendix
\section{General solution of the Riccati equation}
We consider the Riccati equation in the general form as a first-order nonlinear differential equation,\cite{Carinena}
\begin{align}
\frac{d y}{dx} = A(x) y^{2} + B(x) y +C(x).
\end{align}
If we have a particular solution $y = f(x)$, we can then obtain a general solution as $y = f(x) + 1/u$.
The differential equation for $u$ is a linear equation,
\begin{align}
\frac{d u}{d x} &= - \left( 2 A(x) f(x) + B(x) \right) u - A(x).
\end{align}
In terms of the initial value $u(x_{0})$ at $x_{0}$, the solution can be expressed as 
\begin{align}
u(x) &= -\left( \int_{x_{0}}^{x}dx' A(x') e^{-K(x')} \right) e^{ K(x)} + u(x_{0}),
\end{align}
with 
\begin{align}
K(x) &= -\int_{x_{0}}^{x} dx' \left(2 A(x') f(x') + B(x') \right).
\end{align}
Hence we have the general solution as
\begin{align}
y &= f(x) + \frac{1}{ -\left( \int_{x_{0}}^{x}dx' A(x') e^{-K(x')} \right) e^{ K(x)} + u(x_{0})}\,. \label{eq:y}
\end{align}
This solution is well known in mathematics.\cite{Eschrig}
\section{Stability of the Riccati equations}
We now discuss the stability of integrating the Riccati equations with the use of the analytical solutions. 
\subsection{Bulk}
The solution for a homogeneous bulk system is
\begin{align}
a &= \frac{-  \omega_{n} + \sqrt{|\Delta|^{2} + \omega_{n}^{2}}}{\Delta^{\ast}}\,.
\end{align}
Therefore, ${\rm Re} (a \Delta^{\ast}) > 0$ when $\epsilon < \Delta$.
\subsection{Near a vortex}
Near a vortex, we can use the Kramer-Pesch approximation (KPA).\cite{KramerPesch,NagaiPRL,NagaiJPSJ,NagaiOrganic} 
The KPA can be thought of as adding a perturbation to the quasiparticle energy as well as the imaginary part of the order parameter in the Riccati formalism. 
Introducing the variables,
\begin{align}
a &= \bar{a} e^{i \theta}, \\
b &= \bar{b} e^{-i \theta}, \\
\Delta &= \bar{\Delta} e^{i \theta}, 
\end{align}
the Riccati equations can be rewritten as 
\begin{align}
v_{\rm F} \frac{\partial }{\partial s} \bar{a} &= -2  \tilde{\omega}_{n} \bar{a} - \bar{a}^{2} \bar{\Delta}^{\ast} + \bar{\Delta}\,, \\
v_{\rm F} \frac{\partial }{\partial s} \bar{b} &= 2  \tilde{\omega}_{n} \bar{b} + \bar{b}^{2} \bar{\Delta}- \bar{\Delta}^{\ast}. 
\end{align}
The two-dimensional polar coordinates are denoted here as
\begin{align}
\Vec{r} &= (s,y) = r(\cos \theta,\sin \theta).
\end{align}
In these coordinates, $\bar{\Delta}$ reduces to
\begin{align}
\bar{\Delta}(\Vec{r},\Vec{k}_{\rm F}) &= f(\Vec{r}) \Delta_{0} d(\Vec{k}_{\rm F}) \frac{s + i y}{\sqrt{s^{2}+y^{2}}}\,.
\end{align}
By means of KPA, we have
\begin{align}
\bar{a}(\Vec{r},\Vec{k}_{\rm F}) &\sim a_{0}(\Vec{k}_{\rm F}) + a_{1}(\Vec{r},\Vec{k}_{\rm F}),
\end{align}
where
\begin{align}
a_{0}(\Vec{k}_{\rm F}) &= - {\rm sgn}\: [d(\Vec{k}_{\rm F})]\,, \\
a_{1}(\Vec{r},\Vec{k}_{\rm F}) &= -2  \frac{e^{u(\Vec{r})}}{v_{\rm F}} \int_{-\infty}^{s}
\left[ a_{0}(\Vec{k}_{\rm F}) \omega_{n} - i {\rm Im} \bar{\Delta}(\Vec{r}')
\right] e^{- u(\Vec{r}')}ds'.
\end{align}
Here,
\begin{align}
u(\Vec{r}) &= \frac{2}{v_{\rm F}} a_{0}(\Vec{k}_{\rm F}) \int_{0}^{s} {\rm Re} \bar{\Delta}(\Vec{r}') ds'.
\end{align}
The condition ${\rm Re} (\Delta^{\ast} a) > 0$ then translates to
\begin{align}
{\rm Re} \bar{\Delta}^{\ast} \bar{a} &= D(s) \left[  -s 
+ \frac{2 e^{u(\Vec{r})}}{v_{\rm F}} \left( 
s \omega_{n} C(s) - y^{2} E(s) 
\right) \right] > 0\,,
\end{align}
with 
\begin{align}
C(s) &= \int_{-\infty}^{s}e^{- u(\Vec{r}')} ds', \\
D(s) &= \frac{f(r) \Delta_{0} |d(\Vec{k})|}{\sqrt{s^{2}+y^{2}}}\,,\\
E(s) &= \int_{-\infty}^{s} D(s') e^{-u(\Vec{r}')} ds'.
\end{align}
Since $e^{u(\Vec{r})}$ is a localized function at $s = 0$ and the applicable range for the perturbation is $|a_{0}| > |a_{1}|$, 
the condition ${\rm Re} (\Delta^{\ast} a) > 0$ is satisfied in the region $s < 0$ for $\omega_{n} > 0$. 
This means that one can obtain numerically stable solutions in a system containing a vortex. 
Furthermore, as $K(s)$ is a function obtained by integration of ${\rm Re} (\Delta^{\ast}a)$, 
it is an increasing function with $s$ close to and far away from a vortex.
Thus it can result in numerically stable solutions for a system containing
many vortices, as long as the intervortex distances are sufficiently longer than the coherence length.

\section{Path with specular reflections inside a disk}
We illustrate how to generate a path with specular reflections 
inside a circular disk, from a initial point $(x_{0},y_{0})$ 
with initial angle $\theta$.
The linear path that goes through the point $(x_{0},y_{0})$ with the gradient $a = \tan \theta$ can be written as $y = a (x - x_{0}) + y_{0}$.
We find the point of intersection of this path with the circular boundary, which is given by
$x^{2} + y^{2} = r_{c}^{2}$.
The solutions are 
\begin{align}
x_{\pm} &= \frac{a^{2} x_{0} - a y_{0} \pm D}
{1+a^{2}}\,, \\
y_{\pm} &= a (x_{\pm} - x_{0}) + y_{0}\,,
\end{align}
with 
\begin{align}
D &= \sqrt{r_{c}^{2}+ a^{2} r_{c}^{2} -a^{2} x_{0}^{2} + 2 a x_{0} y_{0} - y_{0}^{2}}\,.
\end{align}
Denoting $(x_{c},y_{c}) = (x_{-},y_{-})$ as the point of intersection, 
we have the path as
\begin{align}
y = a (x - x_{0}) + y_{0}, \: \: \: x_{c} < x < x_{0}\,.
\end{align}
The angle of specular reflection $\theta'$ can be found by simple geometry:
\begin{align}
\theta' &= \theta + 2 \delta \theta\,, \\
\delta \theta &= \left\{ \begin{array}{ll}
\alpha - \theta & (\alpha > 0 ) \\
\pi - \theta + \alpha& (\alpha < 0) \\
\end{array} \right. ,
\end{align}
where
$(x_{c},y_{c}) = r_c(\cos \alpha,\sin \alpha)$ in polar coordinates.
The angle $\theta'$ becomes the new momentum direction $\theta_{i-1}$ after $i-1$ specular reflections.

We find the next segment of the path in the direction $\theta_{i-1}$
by adopting as the initial point for the $i$-th path $(x_{i-1},y_{i-1}) = (x_{c},y_{c})$.
The point of intersection is given by
\begin{align}
x_{\pm} &= \frac{a^{2} x_{i-1} - a y_{i-1} \pm D}
{1+a^{2}}\,, \\
y_{\pm} &= a (x_{\pm} - x_{i-1}) + y_{i-1}\,,
\end{align}
where
\begin{align}
D &= \sqrt{r_{c}^{2}+ a^{2} r_{c}^{2} -a^{2} x_{i-1}^{2} + 2 a x_{i-1} y_{i-1} - y_{i-1}^{2}}\,,
\end{align}
with $a = \tan \theta_{i-1}$.
One of the solutions is equal to $x_{i-1}$ and the other solution is 
the next intersection point. Then we have the $i$-th path as
\begin{align}
y = a (x - x_{i-1}) + y_{i-1}, \: \: \: x_{i-1} < x < x_{i}\,.
\end{align}
This method of constructing paths confined in a finite-size system with specular reflections at the boundary by geometry can easily be generalized to a system of arbitrary shape.


\end{document}